%% file: main.tex
\documentclass[runningheads]{llncs}
\usepackage{graphicx}
\input{tex/preamble}

\usepackage{multirow}
\usepackage{makecell} % For multiline cells
\usepackage{subfig} 
\usepackage{cleveref}

\crefformat{section}{\S#2#1#3}
\crefformat{subsection}{\S#2#1#3}
\crefformat{subsubsection}{\S#2#1#3}

\begin{document}
\title{Incremental Adaptive Attack Synthesis}
% USE THIS FOR CAMERA READY
%\thanks{This material is based on research sponsored by DARPA under agreement
%number FA8750-15-2-0087. The U.S. Government is authorized to reproduce and
%distribute reprints for Governmental purposes notwithstanding any copyright
%notation thereon. The views and conclusions contained herein are those of
%the authors and should not be interpreted as necessarily representing the
%official policies or endorsements, either expressed or implied, of DARPA
%or the U.S. Government}} %

\titlerunning{Attack Synthesis for Strings}

% \author{Seemanta Saha}
% \institute{}

\author{Seemanta Saha, William Eiers, Ismet Burak Kadron, \\
Lucas Bang$^{*}$, Tevfik Bultan}
\authorrunning{S. Saha et al.}
\institute{University of California, Santa Barbara 93106, USA \\
    $^*$Harvey Mudd College, Claremont, CA 91711, USA \\
    \email{\{seemantasaha,weiers,kadron,bultan\}@cs.ucsb.edu \\
    $^*$bang@cs.hmc.edu}}

\maketitle  % typeset the header of the contribution

\begin{abstract}
\input{tex/abstract.tex}


\end{abstract}

\section{Introduction}
\label{sec:intro}
\input{tex/intro}

\section{Motivation}
\label{sec:motivation}
\input{tex/motivation}

% \section{Overview}
% \label{sec:overview}
% \input{tex/overview}

% \section{Background} 
% \label{sec:background}
% \input{tex/background}

\section{Synthesizing Adaptive Attacks} 
\label{sec:attacker_model}
\input{tex/attacker-model}

\section{Incremental Attack Synthesis} 
\label{sec:model_counting}
\input{tex/model-counting}
\section{Attack Synthesis Heuristics} 
\label{sec:attack_synthesis_heuristics}
\input{tex/attack_synthesis_heuristics}

% \section{Implementation} 
% \label{sec:implementation}
% \input{tex/implementation}

\section{Implementations and Experiments} 
\label{sec:experiments}
\input{tex/implementation}
\input{tex/experiments}

\section{Case Studies} 
\label{sec:case_studies}
\input{tex/case-studies}

\section{Related Work} 
\label{sec:related_work}
\input{tex/related}

\section{Conclusion} 
\label{sec:conclusion}
\input{tex/conclusion}

% \section*{Acknowledgment}

% \ref{tab:experiment1}\newpage

% Possible bug with bibtex. 
% You may need to
% 1. switch style to plain
% 2. pdflatex, bibtex, pdflatex
% 3. switch style back to template style
% 4. pdflatex

\bibliographystyle{plain}
\bibliography{biblio}

\end{document}

%% file: tex/preamble.tex
\usepackage{cite}
\usepackage{amsmath,amssymb,amsfonts}
\usepackage{textcomp}

\usepackage{graphicx}
\usepackage{color}
\usepackage{listings}
\usepackage{tikz}
\usepackage{pgfplots}
\usepackage{wrapfig}
\usepackage{mdframed}
\usepackage{enumitem}
\usepackage{url}

\usepackage{dcolumn}

\usepackage{booktabs} % For formal tables
\usepackage{subfig}
\usepackage[mathscr]{euscript}

\usepackage{stmaryrd}
\usepackage{mathtools}

\usepackage{enumitem}

\DeclarePairedDelimiterX{\infdivx}[2]{(}{)}{%
  #1\;\delimsize\|\;#2%
}

\usepackage{amssymb}
\usepackage{fancyvrb}
\usepackage{graphicx}

\usepackage{tikz-qtree}

\usetikzlibrary{pgfplots.groupplots}
\usetikzlibrary{fit,shapes,arrows,automata,positioning,trees,shapes,calc}
\usetikzlibrary{plotmarks}
\usetikzlibrary{arrows.meta}

\usepackage[Algorithm,ruled]{algorithm}
\usepackage[noend]{algpseudocode}

\makeatletter
\def\BState{\State\hskip-\ALG@thistlm}
\makeatother

\renewcommand{\floatpagefraction}{.95}

\usepackage{mathtools}

\usetikzlibrary{positioning} 
\tikzstyle{line}=[draw]

\tikzstyle{component} = [rectangle, rounded corners, minimum height=1cm,text centered, draw=black, fill=white!30]

\tikzstyle{input} = [rectangle, rounded corners, minimum width=3cm, minimum height=1cm,text centered, draw=white, fill=white!30]

\tikzstyle{dummy} = [rectangle, rounded corners, text centered, draw=white, fill=white!30]

\tikzstyle{arrow} = [thick,->,>=stealth]

\usepackage[nounderscore]{syntax} % grammar environment
\newcommand{\mop}[1]{\mathrm{#1}} %
\newcommand*\colvec[3][]{\begin{matrix}\ifx\relax#1\relax\else#1\\\fi#2\\#3\end{matrix}}

\definecolor{mygreen}{rgb}{0,0.6,0}
\definecolor{mygray}{rgb}{0.5,0.5,0.5}
\definecolor{mymauve}{rgb}{0.58,0,0.82}

\lstdefinestyle{customc}{%
  belowcaptionskip=1\baselineskip,
  breaklines=true,
  frame=L,
  xleftmargin=\parindent,
  language=C,
  showstringspaces=false,
  basicstyle=\footnotesize\ttfamily,
  keywordstyle=\bfseries\color{green!40!black},
  commentstyle=\itshape\color{purple!40!black},
  identifierstyle=\color{blue},
  stringstyle=\color{orange},
}

\newcommand{\sterm}{\gamma} %
\newcommand{\iterm}{\beta}

\DeclareDocumentCommand \ends {} %
{%
  \mop{\it ends}
}%
\newcommand{\charat}{\mop{\it charat}} %
\tikzstyle{decision} = [diamond, draw, text width=4.5em, text badly centered,
inner sep=0pt]
\tikzstyle{block} = [rectangle, draw, fill=blue!20, text width=5em, text
centered, rounded corners, minimum height=4em]
\tikzstyle{line} = [draw, -latex']
\tikzstyle{cloud} = [draw, ellipse,fill=red!20, node distance=3cm, minimum
height=2em]
\tikzstyle{hexagon} = [regular polygon, draw, regular polygon sides=6, text
width=4.2em, inner
sep=0pt]
\tikzstyle{var} = [rectangle, draw, fill=blue!5, text centered, rounded corners,
minimum width=2em, minimum height=2em, node distance=4cm,
execute at begin node={\begin{varwidth}{12em}},
  execute at end node={\end{varwidth}}]
\tikzstyle{func} = [draw, ellipse, node distance=2cm, minimum height=2em]
\tikzstyle{sink} = [draw, ellipse, double, double distance=1pt, node
distance=2cm, minimum height=2em]

\pgfdeclarelayer{background}
\pgfdeclarelayer{foreground}
\pgfsetlayers{background,main,foreground}
\tikzstyle{materia}=[draw, fill=white!20, text width=6.0em, text centered,
minimum height=1.8em]
\tikzstyle{practica} = [materia, text width=8em, minimum width=10em,
minimum height=3em, rounded corners]
\tikzstyle{module} = [materia, text width=8em, minimum width=8em,
minimum height=3em]
\tikzstyle{driver} = [materia, text width=25em,
minimum height=2em, rounded corners]
\tikzstyle{texto} = [above, text width=6em, text centered]
\tikzstyle{linepart} = [draw, thick, color=black!50, -latex', dashed]
\tikzstyle{ur}=[draw, text centered, minimum height=0.01em]

\tikzset{
      invisible/.style={opacity=0},
      visible on/.style={alt=#1{}{invisible}},
      rect on/.style={alt=#1{rect}{}},
      alt/.code args={<#1>#2#3}{%
        \alt<#1>{\pgfkeysalso{#2}}{\pgfkeysalso{#3}} % \pgfkeysalso doesn't change the path
      },
      on slide/.code args={<#1>#2}{%
        \uncover<#1>{\pgfkeysalso{#2}}
      },
      only on slide/.code args={<#1>#2}{%
        \only<#1>{\pgfkeysalso{#2}}
      },
      rect label/.style={label={[anchor=north west, font=\tiny]north west:{\color{black!75}#1}}},
      rect/.style={rectangle, draw, rounded corners, inner sep=2pt},
      arrow/.style={edge from parent/.style={draw,-latex'}},
      left_arrow/.style={edge from parent path={(\tikzparentnode.south west) -- (\tikzchildnode.north)}},
      right_arrow/.style={edge from parent path={(\tikzparentnode.south east) -- (\tikzchildnode.north)}},
      lc_arrow/.style={edge from parent path={(\tikzparentnode.south)+(-0.5em,0pt)  -- (\tikzchildnode.north)}},
      rc_arrow/.style={edge from parent path={(\tikzparentnode.south)+(0.5em,0pt) -- (\tikzchildnode.north)}},
      down_arrow/.style={edge from parent path={(\tikzparentnode.south) -- (\tikzchildnode.north)}},
      growth parent anchor=south,
      level distance=5em,
      level 1/.style={sibling distance=12em},
      level 2/.style={sibling distance=6em}
    }

\tikzstyle{every state}=[style={font=\tiny},fill=none,draw,text=black, inner sep=0pt,minimum size=1.2em]
\tikzset{
  every edge/.append style={font=\tiny},
  every picture/.style={node distance=3em}
  }

%% file: tex/abstract.tex
Information leakage is a significant problem in modern software systems. Information leaks due to side channels are especially hard to detect and analyze. In this paper, we present techniques for automated synthesis of adaptive side-channel attacks that recover secret values. Our attack synthesis techniques iteratively generate inputs which, when fed to code that accesses the secret, reveal partial information about the secret based on the side-channel observations, reducing the remaining uncertainty about the secret in each attack step. Our approach is incremental, reusing results from prior iterations in each attack step to improve the efficiency of attack synthesis. We use symbolic execution to extract path constraints, automata-based model counting to estimate probabilities of execution paths, and meta-heuristics to maximize information gain based on entropy in order to minimize the number of synthesized attack steps. 

%% file: tex/intro.tex
%!TEX root = ../main.tex
It is common for modern software systems store and manipulate sensitive data. It is crucial for software developers to write code in a manner that prevents disclosure of sensitive data that should be kept secret to unauthorized users. However, computation that occurs over sensitive data can have measurable non-functional characteristics that can reveal information. This can allow a malicious user to infer information about sensitive data by measuring characteristics such as execution time, memory usage, or network delay. This type of unintended leakage of information about sensitive data due to non-functional behavior of a program is called a side-channel vulnerability. In this paper, we focus on automatically synthesizing adaptive side-channel attacks against functions that manipulate secret values. Our attack consists of a sequence of public inputs that a malicious user can use to leak information about a secret by observing side-channel behavior. By synthesizing an attack, we provide an exploit demonstrating the side-channel vulnerability of the function. 

% For a given function $F$ that performs computation over secret values unknown to users of that function, we synthesize a side-channel attack against $F$.
% If our approach fails to find an attack, this can increase
% our confidence that the function is safe against timing-channel attacks.

Our approach uses symbolic execution to extract constraints that characterize the relationship between the secret values in the program, attacker controlled inputs, and side-channel observations. We investigate and compare several methods for selecting the next attack input based on meta-heuristics for maximizing the amount of information gained and automata-based techniques for constraint solving and model counting. Our attack synthesis approach is adaptive, taking into account the information learned about the secret in previous steps while choosing the public input for the next attack step; and it is incremental, re-using the results from prior iterations in order to improve the performance of each attack synthesis step. Moreover, our attack synthesis approach can handle unbounded string constraints in addition to linear arithmetic constraints.

\vskip 1mm
\noindent \textit{Contribution.} Our contributions in this paper can be summarized as follows:
(1) to the best of our knowledge, our work is the first that can automatically synthesize side-channel attacks for programs that manipulate both unbounded string and numeric values; (2) we use a greedy approach that maximizes the information gain about the secret in each attack step and
we use meta-heuristics for searching the input space during attack synthesis, resulting in a generalized attack synthesis approach; (3) we present an incremental attack synthesis approach based on incremental automata-based model counting that reuses the results from prior attack steps in order to improve the efficiency of attack synthesis; and (4) we present experiments on Java functions and case studies demonstrating realistic attack scenarios, and our experiments demonstrate that our attack synthesis approach is effective and our incremental approach drastically improves the performance of attack synthesis, reducing the attack synthesis time by an order of magnitude. 

%In order to support We have added incremental solving to model-counting constraint solver ABC~\cite{ABB15}.

%% file: tex/motivation.tex
\noindent \textit{Motivating Example 1.}
Consider a PIN-based authentication function (Fig.~\ref{fig:check1}) with 
inputs: 
1) a secret PIN $h$, and 
2) a user input, $l$.
Both $h$ and $l$ are strings of digit characters (``0''--``9'') of length $4$. 
We have adopted the nomenclature used in
security literature where $h$ denotes the {\em high-security}
value (the secret PIN) and $l$ denotes the {\em low-security} input value, (the
input that the function compares with the PIN). 
The function compares the PIN and the user input
character by character and returns \texttt{false} as soon as it finds a
mismatch. Otherwise it returns \texttt{true}.

\begin{figure}[!t]
\centering
\begin{minipage}{0.65\linewidth}
\begin{lstlisting}[stepnumber=0,basicstyle=\scriptsize]
public Boolean checkPIN(String h, String l){
  for (int i = 0; i < 4; i++)
    if (h.charAt(i) != l.charAt(i)) 
      return false;
  return true;
}
\end{lstlisting}
\end{minipage}
\vspace{-4mm}
\caption{PIN checking example.}
\label{fig:check1}
\vspace{-6mm}
\end{figure}

This function has a timing side-channel and one can infer information about the secret $h$ by measuring the execution time. 
In this implementation of \texttt{checkPIN} each length of the common prefix
of $h$ and $l$, the number of bytecode instructions that will be executed will differ which may cause an observable difference in execution time.
Notice that if $h$ and $l$ have no common prefix, then \texttt{checkPIN}
will have the shortest execution since the loop body will be executed
only once; this corresponds to execution of 63 Java bytecode
instructions. 
If $h$ and $l$ have a common prefix of one character, 
we see a longer execution since the loop body executes
twice (78 instructions). 
In the case that $h$ and $l$ match completely, \texttt{checkPIN}
has the longest execution (123 instructions).

If we assume that differences in the number of bytecode instructions are observable by measuring the execution time, then there are $5$ observable values since there are $5$ execution paths with different lengths, proportional to the length of the common prefix of $h$
and $l$. In general, using the number of executed bytecode instructions as a measurable
observation can result in observations that are indistinguishable in
practice. Thus, we combine observations into indistinguishability intervals $o \pm \delta$ using
an observability threshold $\delta$. For this example assume that differences among execution path lengths are above this threshold. 

Given this side-channel, an attacker can choose an input and use
the timing observation to determine how much of a prefix of the input has matched the secret.
In order to automate this process, our approach starts with
automatically generating the path constraints and
corresponding execution costs (in terms of number of executed bytecode instructions) using symbolic execution (Table~\ref{tab:example-pcs}). It merges path constraints based on the 
observability threshold, resulting in a set of observability constraints. 
It then uses these constraints to synthesize an attack which
determines the value of the secret PIN. We make use of an uncertainty
function, based on Shannon entropy, to measure the progress of an attack
(Section~\ref{sec:objective_function}). Intuitively, the attacker's
uncertainty, $\mathcal{H}$ starts off at some positive value corresponding to the
initial uncertainty of the sercret, and decreases
during  the attack. When $\mathcal{H} = 0$, the attacker has fully learned
the secret (Table~\ref{tab:example-attack}).

\vskip 2mm

\noindent
\begin{minipage}[!h]{0.45\textwidth}
\input{tables/example-pcs-table}
\end{minipage}
~~\begin{minipage}[!tb]{0.5\textwidth}
\input{tables/example-attack-table}
\end{minipage} 

\vskip 2mm

Suppose that the secret is ``1337''. 
The initial uncertainty
is $\log_2 10^4 = 13.13$ bits of information (assuming uniform distribution).
Our attack synthesizer generates input ``8229'' at the first step and makes
an observation with cost 63, which corresponds to $\psi_1$. This
indicates that ${\charat}(h,0) \not = 8$. 
Similarly, a second synthesized input, ``0002'', implies $\charat(h,0) \not = 0$ and the uncertainty is again reduced.
At the third step the synthesized input
``1058'' yields an observation of cost 78. 
Hence, $\psi_2$ is the correct path constraint 
to update our constraints on $h$, which becomes

\vskip 1mm

$\charat(h,0) \not = 8 \land \charat(h,0) \not = 0 \land \charat(h,0) = 1 \land \charat(h,1) \not = 0$ 

\vskip 1mm

We continue synthesizing inputs and updating the constraints on $h$,
which tell us more information about $h$, until the
secret is known after 27 steps. At the final step,
we make an observation which corresponds to $\psi_5$ indicating a full
match and the remaining uncertainty is 0. As in this example,
the goal of our search for attack inputs is to drive the entropy that characterizes the remaining
uncertainty about the secret to 0.
Thus, we propose entropy optimization techniques. This particular type of attack is 
called a \emph{segment attack} which is known to be a serious source of security
flaws~\cite{BAP16,CRIMEattack,Kel02,Law09.2,Nel10}, and it is exponentially shorter than a brute-force attack. Our approach automatically synthesizes a segment attack.

\vskip 1mm
\noindent \textit{Motivating Example 2.} Consider another example (Figure~\ref{fig:str_compare}). If secret value $h$ is lexicographically smaller than user input $l$, the execution time of \textit{stringInequality} corresponds to 47 instructions, and 62 otherwise. Symbolically executing the \textit{stringInequality} method (note that, we do not symbolically execute the \textit{compareTo} method from Java's string library but capture it as a string constraint directly), two path constraints are inferred with distinguishable observations shown in Table~\ref{tab:example-2-pcs}. For simplicity, consider the secret domain to be from ``AA" to ``ZZ" ($26^{2} = 676$ strings), the secret value is ``LL" and the first attack input is ``AA". In Table~\ref{tab:example-2-attack-non-optimal} we show an attack that recovers the secret in 20 attack steps. %Java bytecode 

%For the \texttt{checkPIN} example, at any attack step, any input  that satisfies the constraint for the next attack interaction will leak the same amount of information. Hence, choosing any of the inputs as attack input will be an optimal step for the attacker. For this kind of attack (\emph{segment attack}) synthesis, our Model-based (M) approach would be enough to generate an optimal attack (see~\cref{sec:attack_synthesis_heuristics}). 
%It is not generally the case that choosing any constraint-satisfying input can generate an optimal attack. 

%One approach to generating an attack is to find a satisfying solution (i.e., model) to
%the constraints on the low variable that is consistent with the observations about the
%secret we have accumulated so far. We also have to make sure that we do not repeat the public input values that we have already tried. 

We can generate an attack like the one shown in Table~\ref{tab:example-2-attack-non-optimal}
by finding a satisfying solution (i.e., model) to the constraints on the low variable that is consistent with the observations about the secret we have accumulated so far. We call this the Model-based (M) approach (see~\cref{sec:attack_synthesis_heuristics}), and this approach does generate optimal segment attacks as we discussed above for the example shown in Fig.~\ref{fig:check1}. However, for the example shown in Figure~\ref{fig:str_compare} the Model-based approach cannot generate an optimal attack.  

%In this case, the input attacker tries at each attack step is not guaranteed to be an optimal input (maximizing information leakage). 

%A similar result occurs with our Model Based (M) approach, as the attack input we try is  randomly chosen from a uniform distribution over the constraint-satisfying inputs, not an optimal input. Note that, you can be lucky and generate the attack with few steps but this not the average case. On the contrary, instead of getting one input randomly 

The attack shown in Table~\ref{tab:example-2-attack-non-optimal} recovers the secret
but it is not optimal in terms of the length of the attack. 
In order to generate an optimal attack we have to choose an input that maximizes the amount of information leaked in each attack step. Then, we can generate the attack shown in Table~\ref{tab:example-2-attack-optimal} which is optimal and requires only 9 steps. This corresponds to a binary search, finding the middle point to divide the domain of secret value in a balanced way. For our example, the domain size $d$ is $26^{2}$ and taking $\log_2 d$, we get $\lceil 9.40 \rceil = 10$ attack steps in the worst case. In order to generate the optimal attack automatically, we need to construct an objective function  (see~\cref{sec:objective_function}) characterizing the information gain for each attack step and use optimization techniques (see~\cref{sec:attack_synthesis_heuristics}) to maximize the objective function.

Let us have a look at the constraints on secret value $h$ at each attack step for the optimal attack from Table~\ref{tab:inc-constraints-table}. At each attack step we gain new information about the secret value $h$ and a new constraint is added to the already existing constraint $C_h$. The constraint $C_h$ grows and becomes more complex in each attack step. Constraint solving and model counting are the most expensive parts of our approach. So, if we can reuse prior solutions to constraint solving and model counting to take advantage of the incremental nature of attack synthesis, we can increase the efficiency of our approach. We call this approach \textit{incremental attack synthesis} (see~\cref{sec:model_counting}) and demonstrate that it improves the efficiency of attack synthesis significantly (see~\cref{sec:experiments}).

\begin{figure}[!t]
\centering
\begin{minipage}{0.6\linewidth}
\begin{lstlisting}[stepnumber=0,basicstyle=\scriptsize]
public static void stringInequality(String h, String l) {
    if(h.compareTo(l) <= 0) {
      for (int i = 1; i > 0 ; i--);
    } else {
      for (int i = 5; i > 0 ; i--);
    }
}
\end{lstlisting}
\end{minipage}
\vspace{-4mm}
\caption{String inequality example.}
\label{fig:str_compare}
\vspace{-6mm}
\end{figure}

\vskip 2mm
\noindent
\begin{minipage}[!h]{0.22\textwidth}
\input{tables/example-2-pcs-table}
\end{minipage}
~\begin{minipage}[!h]{0.46\textwidth}
\input{tables/example-2-nonoptimal-attack-table}
\end{minipage}
\begin{minipage}[!h]{0.30\textwidth}
\input{tables/example-2-optimal-attack-table}
\end{minipage}
\vskip 2mm

\vskip 2mm
\noindent
\begin{minipage}[!h]{1.0\textwidth}
\input{tables/inc-constraints-table.tex}
\end{minipage}
\vskip 2mm

% In addition, we extended
% an existing  solver and model counter for string and integer constraints
% called ABC~\cite{ABB15} for performing incremental string analysis, which
% greatly increases the efficiency of our attack synthesis approach. 

%% file: tables/example-pcs-table.tex
% \begin{table}[!ht]
\centering
\captionof{table}{Observation constraints generated by symbolic execution of the function in Figure~\ref{fig:check1}.}
\label{tab:example-pcs}
\begin{scriptsize}
\begin{tabular}{|l||l|l|}
\hline
$i$ & Observation Constraint, $\psi_{i}$                 & $o$       \\ \hline \hline
1   & $\charat(l,0)  \not = \charat(h,0)$         & 63      \\ \hline
2   & $\charat(l, 0)      = \charat(h, 0) \land$  & 78      \\
    & $\charat(l, 1) \not = \charat(h, 1)$        &         \\ \hline
3   & $\charat(l,0)       = \charat(h, 0)\land$   & 93      \\
    & $\charat(l,1)       = \charat(h, 1)\land$   &         \\
    & $\charat(l,2)  \not = \charat(h, 2)$        &         \\ \hline
4   & $\charat(l,0)       = \charat(h, 0)\land$   & 108     \\
    & $\charat(l,1)       = \charat(h, 1)\land$   &         \\
    & $\charat(l,2)       = \charat(h, 2)\land$   &         \\
    & $\charat(l,3)  \not = \charat(h, 3)$        &         \\ \hline
5   & $\charat(l,0)       = \charat(h, 0)\land$   & 123     \\
    & $\charat(l,1)       = \charat(h, 1)\land$   &         \\
    & $\charat(l,2)       = \charat(h, 2)\land$   &         \\
    & $\charat(l,3)       = \charat(h, 3)$        &         \\ \hline
\end{tabular}
\end{scriptsize}
% \end{table}

%% file: tables/example-attack-table.tex
% \begin{table}[!h]
\centering
\captionof{table}{Attack inputs ($l$), uncertainty about the secret ($\mathcal{H}$), and observations ($o$). 
Prefix matches are shown in \textbf{bold}. }
\label{tab:example-attack}
\begin{scriptsize}
\begin{tabular}{|r|r|r|r|||r|r|r|r|}
\hline
Step & $\mathcal{H}$ & $l$ & $o$ & Step & $\mathcal{H}$ & $l$ & $o$ \\ \hline \hline
1  & 13.13 & ``\textbf{}8299'' & 63 &       15 & 5.906 & ``\textbf{13}92'' & 93  \\
2  & 12.96 & ``\textbf{}0002'' & 63 &       16 & 5.643 & ``\textbf{13}16'' & 93  \\
3  & 9.813 & ``\textbf{1}058'' & 78 &       17 & 5.321 & ``\textbf{13}08'' & 93  \\
4  & 9.643 & ``\textbf{1}477'' & 78 &       18 & 4.906 & ``\textbf{13}62'' & 93  \\
5  & 9.451 & ``\textbf{1}583'' & 78 &       19 & 4.321 & ``\textbf{13}78'' & 93  \\
6  & 9.228 & ``\textbf{1}164'' & 78 &       20 & 3.169 & ``\textbf{133}8'' & 108 \\
7  & 8.965 & ``\textbf{1}950'' & 78 &       21 & 3.000 & ``\textbf{133}2'' & 108 \\
8  & 8.643 & ``\textbf{1}220'' & 78 &       22 & 2.807 & ``\textbf{133}4'' & 108 \\
9  & 8.228 & ``\textbf{1}786'' & 78 &       23 & 2.584 & ``\textbf{133}3'' & 108 \\
10 & 7.643 & ``\textbf{1}817'' & 78 &       24 & 2.321 & ``\textbf{133}0'' & 108 \\
11 & 6.643 & ``\textbf{1}664'' & 78 &       25 & 2.000 & ``\textbf{133}5'' & 108 \\
12 & 6.491 & ``\textbf{13}42'' & 93 &       26 & 1.584 & ``\textbf{133}6'' & 108 \\
13 & 6.321 & ``\textbf{13}28'' & 93 &       27 & 0.000 & ``\textbf{1337}'' & 123 \\
14 & 6.129 & ``\textbf{13}86'' & 93 &          &       &      &   \\ \hline
\end{tabular}
\end{scriptsize}
% \end{table}

%% file: tables/example-2-pcs-table.tex
% \begin{table}[!ht]
\centering
\captionof{table}{Observation constraints of the function in Figure~\ref{fig:str_compare}}
\vspace{-2mm}
\label{tab:example-2-pcs}
\begin{scriptsize}
\begin{tabular}{|l||l|l|}
\hline
$i$ & $\psi_{i}$                 & $o$       \\ \hline \hline
1   & $h \leq l$ & 42 \\ \hline
2   & $h > l$  & 67 \\ \hline
\end{tabular}
\end{scriptsize}
% \end{table}

%% file: tables/example-2-nonoptimal-attack-table.tex
% \begin{table}[!h]
\centering
\captionof{table}{Non-optimal attack}
\vspace{-2mm}
\label{tab:example-2-attack-non-optimal}
\begin{scriptsize}
\begin{tabular}{|r|r|r|r|||r|r|r|r|}
\hline
Step & $\mathcal{H}$ & $l$ & $o$ & Step & $\mathcal{H}$ & $l$ & $o$ \\ \hline
1 & 9.40 & ``AC" & 67 & 11 & 7.56 & ``PJ" & 42 \\
2 & 9.39 & ``AE" & 67 & 12 & 6.82 & ``PI" & 42 \\
3 & 9.39 & ``JZ" & 67 & 13 & 6.80 & ``NA" & 42 \\
4 & 8.70 & ``XE" & 42 & 14 & 5.70 & ``LZ" & 42 \\
5 & 8.41 & ``XB" & 42 & 15 & 4.64 & ``LI" & 67 \\
6 & 8.40 & ``KQ" & 67 & 16 & 4.00 & ``LR" & 42 \\
7 & 8.33 & ``XA" & 42 & 17 & 3.00 & ``LK" & 67 \\
8 & 8.32 & ``KU" & 67 & 18 & 2.58 & ``LO" & 42 \\
9 & 8.30 & ``SI" & 42 & 19 & 1.58 & ``LM" & 42 \\
10 & 7.60 & ``KZ" & 67 & 20 & 0.00 & ``LL" & 42 \\ \hline
\end{tabular}
\end{scriptsize}
% \end{table}

%% file: tables/example-2-optimal-attack-table.tex
% \begin{table}[!h]
\centering
\captionof{table}{Optimal attack}
\label{tab:example-2-attack-optimal}
\vspace{-2mm}
\begin{scriptsize}
\begin{tabular}{|r|r|r|r|}
\hline
Step & $\mathcal{H}$ & $l$ & $o$ \\ \hline
1 & 8.40 & ``MZ" & 42 \\
2 & 7.40 & ``GM" & 67 \\
3 & 6.40 & ``JS" & 67 \\
4 & 5.43 & ``LI" & 67 \\
5 & 4.39 & ``MD" & 42 \\
6 & 3.32 & ``LS" & 67 \\
7 & 2.32 & ``LN" & 67 \\
8 & 1.00 & ``LK" & 67 \\
9 & 0.00 & ``LL" & 42 \\ \hline
\end{tabular}
\end{scriptsize}
% \end{table}

%% file: tables/inc-constraints-table.tex
% \begin{table}[!ht]
\centering
\captionof{table}{Incremental nature of constraints at each step of adaptive attack.}
\vspace{-2mm}
\label{tab:inc-constraints-table}
\begin{scriptsize}
\begin{tabular}{|l|l|l|}
\hline
Attack step & Attack input & Constraint on secret value, $C_h$  \\ \hline \hline
1 & ``MZ" & $h <= ``MZ"$ \\ \hline
2 & ``GM" & $h <= ``MZ" \land h > ``GM"$ \\ \hline
3 & ``JS" & $h <= ``MZ" \land h > ``GM" \land h > ``JS"$ \\ \hline
... & ... & ... \\ \hline
9 & ``LL" & \makecell{$h <= ``MZ" \land h > ``GM" \land h > ``JS" \land$ \\ $h > ``LI" \land h <= ``MD" \land h <= ``LS" \land$ \\ $h > ``LN"  \land h > ``LK" \land h <= ``LL"$} \\ \hline
\end{tabular}
\end{scriptsize}
% \end{table}

%% file: tex/attacker-model.tex
%\vspace{-15mm}
We use a two-phase attack synthesis approach as shown in Fig.~\ref{figure:esorics-overview} and Algorithm~\ref{proc:attacker}. We consider a function $F$ that takes as input a secret $h \in \mathbb{H}$ and an attacker-controlled input $l \in \mathbb{L}$ and that generates side-channel observations $o \in \mathbb{O}$. 
%\vskip 1mm

\noindent \textit{Static Analysis Phase.} 
In the first phase we generate observation constraints from $F$ as shown in Algorithm~\ref{proc:gen-constraints}.
First, we perform symbolic execution on $F$ with the
secret ($h$) and the attacker controlled input ($l$) marked as 
symbolic~\cite{King:1976:SEP:360248.360252,Pasareanu:2013:ASE}.
Symbolic execution runs $F$ on symbolic rather than concrete inputs
resulting in a set of path constraints $\Phi$. Each $\phi \in \Phi$ is a logical formula that characterizes the set of inputs that execute some path in $F$.
During symbolic execution, we keep track of a side-channel observation 
for each path. For timing side-channels,  
as in other works in this area, we model the
execution time of the function by the number of instructions
executed~\cite{BAP16,Pasareanu:2016:MSC,PhanCSF2017}. We assume
that the observable values are noiseless, i.e., multiple executions of the
program with the same input value will result in the same observable value. 
We augmented symbolic execution to return a function that maps
a path constraint $\phi$ to an observation $o$. 
Since an attacker cannot extract information from program paths that have
indistinguishable side-channel observations, 
we combine observationally similar path constraints via disjunction (Algorithm~\ref{proc:gen-constraints}, line 4), where
we say that $o$ and $o'$ are in the same equivalence class ($o \sim o'$) if and only if $|o - o'| < \delta$. The resulting
\textit{observation constraints} (denoted $\psi_o$ and $\Psi$) characterize the relationship
between the secret ($h$) the attacker input ($l$) and indistinguishable side-channel observations ($o$).  
\vskip -10mm
\begin{minipage}[t]{0.9\textwidth}
\begin{figure}[H]
   \includegraphics[width=\linewidth]{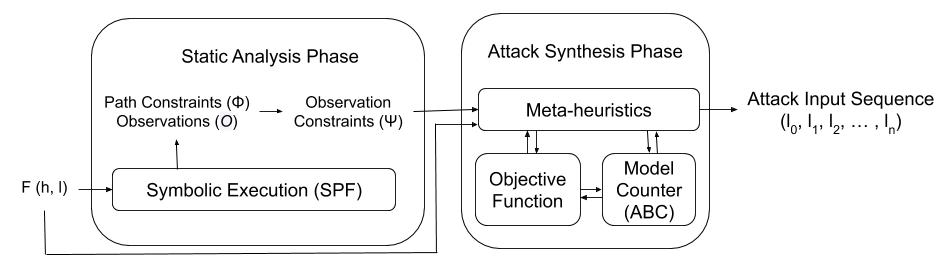}
   \vskip -3mm
   \caption{Overview of Attack Synthesis Approach}
   \label{figure:esorics-overview}
   %\vspace*{-1em}
\end{figure}
\end{minipage}
\vskip -4mm
\begin{algorithm}
\begin{small}
\caption{\textsc{SynthesizeAttack}($F(h,l), C_h, h^*$) \\
This algorithm calls the \textsc{GenerateConstraints} and \textsc{RunAttack} functions 
to synthesize adaptive attacks.
}
\label{proc:attacker}
\begin{algorithmic}[1]
\State $\Psi \leftarrow \textsc{GenerateConstraints}(F(h,l))$

\State \textsc{RunAttack}$(F(h,l), \Psi, C_h, h^*)$

\end{algorithmic}
\end{small}
\end{algorithm} 
\vskip -4mm
\noindent \textit{Attack Synthesis Phase.} 
The second phase synthesizes a sequence of inputs that allow an attacker
to adaptively learn the secret (Algorithm~\ref{proc:run-attack}). During this phase, we fix a secret
$h^*$, unknown to the attacker.
We maintain a constraint $C_h$ on the possible values of the secret $h$. 
Initially, $C_h$ merely specifies the domain of the secret.
We call algorithm \textsc{AttackInput-SA}, 
which uses the simulated annealing technique to maximize information gain about the secret expressed as entropy (as discussed in~\cref{sec:attack_synthesis_heuristics}), to determine the input value $l^*$ for the current attack step. Then, the observation $o$ that corresponds to running the program under attack with $h^*$ and $l^*$ is revealed by running the function using the public input $l^*$. We update $C_h$ to reflect the new constraint on $h$ implied by the attack input and observation---we instantiate the corresponding observation constraint, $\psi_o[l \mapsto l^*]$, and conjoin it with the current $C_h$ (line 5). Based on $C_h$, we compute an uncertainty measure for $h$ 
at every step using Shannon entropy~\cite{shannon48,Cover2006}, denoted $\mathcal{H}$ (Section~\ref{sec:objective_function}). 
The goal is to generate inputs which drive $\mathcal{H}$ as close as possible to zero, in which case there is no uncertainty left about the secret and the secret is fully known.
This attack synthesis phase is repeated until it is not possible to reduce the uncertainty, $\mathcal{H}$, any further.
\vskip -4mm
\begin{algorithm}
% \begin{scriptsize}
\caption{\textsc{GenerateConstraints}($F(h,l)$) \\
Performs symbolic execution on function $F$ with secret string $h$ and attacker-controlled string $l$. The resulting path constraints are combined according to indistinguishability of observations.
}
\label{proc:gen-constraints}
\begin{algorithmic}[1]

% \State $h_{sym} \leftarrow$ \textsc{SymbolicString}(``h''), $l_{sym} \leftarrow$ \textsc{SymbolicString}(``l'')

\State $\Psi \leftarrow \emptyset$

\State $(\Phi, \mathcal{O}, \textit{obs}) \leftarrow$ \textsc{SymbolicExecution}$(F(h, l))$ 

\For{$o \in \mathcal{O}$}

\State  $\psi_o \leftarrow \bigvee_{\phi \in \Phi : \textit{obs}(\phi) \sim o} \phi$

\State $\Psi \leftarrow \Psi \cup \{ \psi_o \}$

\EndFor

\State \Return $\Psi$

\end{algorithmic}
% \end{scriptsize}
\end{algorithm} 
\vskip -12mm
\begin{algorithm}
% \begin{small}
\caption{\textsc{RunAttack}($F(h,l), \Psi, C_h, h^*$) \\
Synthesizes a sequence of attack inputs, $l^*$, for $F(h,l)$, given observation constraints $\Psi$, initial constraints on $h$ ($C_h$), and unknown secret $h^*$.}
\label{proc:run-attack}
\begin{algorithmic}[1]

% \State $U \leftarrrow \mathcal{H}(C_h)$
\State $\mathcal{H} \leftarrow \textsc{Entropy}(C_h)$

\While{$\mathcal{H} > 0$}

\State $l^* \leftarrow \textsc{AttackInput-SA}(C_h, \Psi)$

\State $o \leftarrow F(h^*,l^*)$

\State $C_h \leftarrow C_h \land \psi_{o}[l \mapsto l^*]$

\State $\mathcal{H} \leftarrow \textsc{Entropy}(C_h)$

% \State $i \leftarrow i + 1$

\EndWhile

\end{algorithmic}
% \end{small}
\end{algorithm} 

%% file: tex/model-counting.tex
%!TEX root = ../main.tex
\vskip -2mm
In this section, we first describe the objective function we use to guide the synthesis of each attack step. Then, we discuss the use of automata-based model counting for computing the objective function. Finally, we describe our incremental approach to attack synthesis that reuses results of model counting queries from prior steps for improving efficiency. 
\vskip -8mm
\subsection{Objective Function for Information Gain}
\vskip -1mm
\label{sec:objective_function}
Here we derive an objective function to measure the amount of information an attacker expects to gain by choosing an input value $l_{val}$ to be used in the attack search heuristics discussed in~\cref{sec:attack_synthesis_heuristics}. In the following discussion, $H$, $L$, and $O$ are random variables representing high-security input, low-security input, and side-channel observation,
respectively. We use entropy-based metrics from the theory of quantitative information flow~\cite{Smi09}. Given probability function
$p(h)$, the \emph{information entropy} of 
$H$, denoted $\mathcal{H}(H)$, which we interpret as the initial \emph{uncertainty} about the secret, is 

\begin{footnotesize}
\begin{equation}
\mathcal{H}(H) = - \sum_{h \in  \mathbb{H}} p(h) \log_2 {p(h)}
\label{eq:entropy}
\end{equation}
\end{footnotesize}

\noindent Given conditional distributions $p(h | o, l)$, and $p(o|l)$, we quantify the attacker's expected \emph{updated uncertainty} about $h$, given a candidate choice of $L = l_{\mathit{val}}$, with the expectation taken over all possible observations, $o \in O$. We compute the  
\emph{conditional entropy of \ $H$ given $O$ with $L = l_{\mathit{val}}$} as

\begin{footnotesize}
\begin{equation}
\mathcal{H}(H|O, L = l_{\mathit{val}}) = 
-\sum_{o \in \mathbb{O}} p(o|l_{\mathit{val}}) \sum_{h \in \mathbb{H}} p(h | o, l_{\mathit{val}}) \log_2 {p( h | o, l_{\mathit{val}})}
\label{eq:condentropy}
\end{equation}
\end{footnotesize}

\noindent Now we can compute the expected amount of information \textit{gained} about 
$h$ by observing
$o$ after running the function $F$ with a specific input $l_{\mathit{val}}$. The \emph{mutual information} between $H$ and $O$, given $L = l_{\mathit{val}}$ denoted $\mathcal{I}(H;O | L = l_{\mathit{val}})$ is the difference between the initial entropy of $H$ and the conditional entropy of $H$ given $O$ when $L = l_{\mathit{val}}$:

\begin{footnotesize}
\begin{equation}
\mathcal{I}(H;O | L = l_{\mathit{val}}) = \mathcal{H}(H) - \mathcal{H}(H | O, L = l_{\mathit{val}})
\label{eq:mutinfo}
\end{equation}
\end{footnotesize}

\vskip -2mm
Equation~(\ref{eq:mutinfo}) serves as our objective function. Providing
input $l_{\mathit{val}} = l^*$ which maximizes $\mathcal{I}(H;O | L = l_{\mathit{val}})$
maximizes information gained about $h$. Equations~(\ref{eq:entropy})
and~(\ref{eq:condentropy}) rely on $p(h)$, $p(o|l)$, and $p(h|o,l)$, which
may change at every step of the attack. Recall that during the attack,
we maintain a constraint on the secret, $C_h$. Assuming that all secrets
that are consistent with $C_h$ are equally likely, at each step, we can
compute the required probabilities using model counting. Given a formula $f$,
performing model counting on $f$ gives the number of satisfying solutions
for $F$, which we denote $\#f$.  Thus, we observe that
$p(h) = 1 / \# C_h$ if 
$h$ satisfies $C_h$ 
and is $0$ otherwise. 
Hence, Equation~\ref{eq:entropy} reduces to 
$\mathcal{H}(H) = \log_2(\# C_h)$.

Algorithm~\ref{proc:gen-constraints} gives us side-channel observations
$\mathcal{O} = \{o_1, \ldots, o_n \}$ 
and constraints over $h$ and $l$ corresponding to each $o_i$, 
$\Psi = \{ \psi_1, \ldots, \psi_n \}$. The probability that the secret has a particular value, constrained by a particular $\psi_i$, for a given $l_{\mathit{val}}$ can be computed by instantiating $\psi_i$ with $l_{\mathit{val}}$ and then model counting. Thus, $p(h|o_i,l_{\mathit{val}}) = 1 / {\# (C_h \land \psi_i)[l \mapsto l_{\mathit{val}}]}$.
Similarly, $p(o_i|l_{\mathit{val}}) = {\# (C_h \land \psi_i)[l \mapsto l_{\mathit{val}}]} / \# C_h[l \mapsto l_{\mathit{val}}]$.

In this paper, the \textsc{Entropy} (Equation~(\ref{eq:entropy})) and \textsc{MutualInfo} (Equation~(\ref{eq:mutinfo})) functions refer to the appropriate entropy-based computation just described, where $p(h)$, $p(o|l)$, and $p(h|o,l)$ are computed using the \textsc{ModelCount} algorithm described in the next section. Using \textsc{MutualInfo}, an attacker can optimize the information gain by trying many different $l_{val}$ values and computing the corresponding \textsc{MutualInfo}. Observe that this process involves model counting for instantiating constraints for many values of $l_{val}$. In the next section we describe how to perform this model counting step efficiently. 
\vspace{-4mm}
\subsection{Automata-Based Constraint Solving and Model Counting}
\vspace{-1mm}
As mentioned above, we compute entropy, which is used in the objective function for information gain, using model counting. For this purpose, we use and extend the Automata-Based Model Counter (ABC) tool, which is a constraint solver for string and numeric constraints with model counting capabilities~\cite{ABB15}. The constraint language for ABC supports all numeric constraints solved by off the shelf constraints solvers as well as typical string operations such
as {\em charAt, length, indexOf, substring, begins, concat,} $<$, $=$, etc. Given a constraint $C$, ABC constructs a  multi-track deterministic finite automaton (DFA) $A_C$ that characterizes all solutions for the constraint $C$, where $\mathcal{L}(A_C)$ corresponds to the set of solutions for $C$. For each string term $\sterm$ or integer term $\iterm$ in the constraint grammar \cite{aydin2018parameterized}, ABC implements an automata constructor function which generates an automaton $A$ that encodes the set of satisfying solutions for the term. Note that variables within string terms and integer terms appear in separate automata, as separate encodings are used for each (ASCII for strings, binary encoding for integers).  ABC implements specialized DFA construction algorithms for atomic string operations. Boolean operations ($\land$, $\lor$, $\neg$) are handled using standard DFA operations (intersection, union, and complement, respectively).

ABC counts the number of models (solutions) for a constraint $C$ by first constructing the corresponding automaton $A_C$ and using the observation that number of strings of length $k$  in $\mathcal{L}(A_C)$ is equal to the number of accepting paths of length $k$ in the DFA  $A_C$. Consequently, ABC treats the DFA $A_C$
that results from solving $C$ as a graph where DFA states are graph vertices and the weight of an edge is the number of symbols that have a transition between the source and destination vertices (states) of that edge.  A dynamic programming algorithm that computes the $k$th power of the adjacency matrix of the  graph is used to count
the number of accepting paths in the DFA of length $k$ (or less than or
equal to $k$)~\cite{ABB15}.

% \begin{figure}[t]
%   \centering
%   \footnotesize
%   \input{figures/grammar}

%   \caption{Constraint language grammar}
%   \label{figure:grammar}
% \end{figure}

% \begin{figure}[]
%   \centering
%   \input{figures/automaton_example}
%   \caption{Automata constructed for Example~\ref{ex1}}
%   \label{figure:result_ex4}
%   \vspace*{-1em}
% \end{figure}
\vspace{-4mm}
\subsection{Incremental Constraint Solving and Model Counting}
% Consider the following scenario: We use ABC to solve and model count a
% constraint $C_h$. Then, later on, we need to solve and model
% count the constraint $C_h \land \psi_o[l \mapsto l^*]$. This is a common scenario during 
% attack synthesis, in which the constraint on the secret
% ($C_h$) is updated by conjoining it with the current instantiated observation constraint ($\psi_o[l \mapsto l^*]$)
% (Algorithm~\ref{proc:run-attack}, line 5). 
% Then, we need to compute the entropy for the updated constraint using 
% model counting (Algorithm~\ref{proc:run-attack}, line 6).
\vspace{-2mm}
Attack synthesis requires solving and model counting the constraint on the secret, $C_h$, and updating it with the current instantiated observation constraint, $\psi_o[l \mapsto l_{val}]$ (Algorithm~\ref{proc:run-attack}, line 5). This results in a new constraint, $C_h \land \psi_o[l \mapsto l_{val}]$, which we then compute the entropy for (Algorithm~\ref{proc:run-attack}, line 6). As this process is executed many times, multiple calls to ABC are required, often with similar constraints. In each iteration ABC starts from scratch re-solving each sub-constraint again and constructing a DFA for each of them, then combining them using DFA intersection. Note that, during attack synthesis, $C_h$ can become
a complex combination of constraints that represent what we learned over the
course of the attack. Then ABC would be unnecessarily re-solving the subconstraints of $C_h$ in each attack step.  To summarize, we observe that,
during attack synthesis 1) the constraint that characterizes the set of
secrets that are consistent with the  observations and low inputs ($C_h$)
is constructed incrementally, and 2) computing entropy using incremental
constraint solving can improve the performance by exploiting the incremental
nature of attack synthesis. 

We implemented incremental constraint solving and model counting by extending ABC so that it retains state over successive calls. Given a constraint, ABC constructs the automaton 
representing the set of solutions to the constraint, which is then stored for use in later
calls. The steps of attack synthesis involve two types of model counting for $C_h$ and $\psi_o[l \mapsto l_{val}]$: 
during \textsc{MutualInfo} when an attacker optimizes the attack by trying many different $l_{\mathit{val}}$, and in
\textsc{Entropy}, during computation of the remaining uncertainty. In both situations, model counting is required on many different constraints, and most of the sub-constraints come from previous iterations. We augmented ABC with an interface so that, given $C_h \land \psi_o[l \mapsto l_{\mathit{val}}]$, we can check if an automaton has already been constructed for either $C_h$ or $\psi_o[l \mapsto l_{val}]$, and if so, to get the already constructed automata for them, rather than re-solving each constraint. Note that for the purposes of model counting, $\psi_o[l \mapsto l_{\mathit{val}}]$ can be represented as $\psi_o \land l = l_{\mathit{val}}$. Our incremental model counting approach is outlined in Algorithm~\ref{proc:inc-cons-solving}. Given the constraint $C_h \land \psi_o \land l=l_{\mathit{val}}$, \textsc{GetDFA} retrieves the previously constructed automaton for $C_h$, $A_{C_h}$. 
%Since the observation constraint $\psi_o$ does not change while synthesizing one step of the attack, $A_{\psi_o}$ is constructed and re-used in subsequent model counting queries (lines 2-5). 
Algorithm~\ref{proc:inc-cons-solving} is called with a new observation constraint $\psi_o$ in each attack step, for which the automaton must first be constructed. Subsequent calls with the same $\psi_o$ use the previously constructed automaton. A new $A_{l=l_{\mathit{val}}}$ must be constructed for each model counting query (as each query involves a different $l_{val}$). The final automaton $A$ is constructed using automata product from $A_{C_h}, A_{\psi_o}, A_{l=l_{\mathit{val}}}$. $A$ is exactly the same automaton constructed from $C_h \land \psi_o \land l=l_{\mathit{val}}$, but it is constructed incrementally, thus allowing re-use of previously constructed automata.

\vskip -4mm
\begin{algorithm}
% \begin{scriptsize}
\caption{\textsc{ModelCountIncremental}($C_h \land \psi_o \land l = l_{\mathit{val}}$) \\
Performs incremental model counting for constraint $C_h \land \psi_o \land l = l_{\mathit{val}}$.
}

\label{proc:inc-cons-solving}
\begin{algorithmic}[1]

\State $A_{C_h} \leftarrow \mbox{\sc GetDFA}(C_h)$\

\If {$\mbox{\sc IsConstructed}(\psi_o)$} \

\State $A_{\psi_o} \leftarrow \mbox{\sc GetDFA}(\psi_o)$\

\Else\

\State $A_{\psi_o} \leftarrow \mbox{\sc Construct}(\psi_o)$\

\EndIf

\State $A_{l = l_{\mathit{val}}} \leftarrow \mbox{\sc Construct}(l = l_{\mathit{val}})$\

\State $A \leftarrow A_{C_h} \cap A_{\psi_o} \cap A_{l = l_{\mathit{val}}}$\

\State \Return $\mbox{\sc ModelCount}(A)$

\end{algorithmic}
% \end{scriptsize}
\end{algorithm}
\vspace{-8mm}

%% file: tex/attack_synthesis_heuristics.tex
%!TEX root = ../main.tex
\vskip -2mm
At every attack step the attacker's goal is to choose a low input $l^*$ that 
reveals information about $h^*$. Here we will describe techniques based on constraint solving and meta heuristics for synthesizing attack inputs $l^*$. Meta heuristic approaches explore a subset of the possible low inputs. In order to search the space efficiently, we first observe that we need to restrict the search to those $l$ that are consistent with $C_h$, which we now discuss.

\vskip 1mm \noindent \textit{Constraint-based Model Generation of Low Inputs.}
The first $l$ value can be chosen arbitrarily since initially we do not have any information about the secret $h$. After the first step, our attack synthesis algorithm maintains a constraint $C_h$ which captures all $h$ values that are consistent with the observations so far (Algorithm~\ref{proc:run-attack}, line 5).
Using the observation constraints $\Psi$ (which identify the relation among the secret $h$, public input $l$ and the observation $o$), we project
$C_h$ to a constraint on the input $l$, which we call $C_l$, and we restrict our search on $l$ to the set of values allowed by $C_l$.  I.e., we
only look for $l$ values that are consistent with what we know about $h$ (which is characterized by $C_h$) with respect to $\Psi$. This approach is implemented in \textsc{GetNeighborInput} function which returns an $l_{\mathit{val}}$ by mutating the previous $l_{\mathit{val}}$.

%These two functions were 
%further classified as Restricted (R), in which only models of $C_l$ are generated, or non-restricted (NR), in which we did not enforce $l_{val}$ to be a model of $C_l$. For 
% \ref{proc:sim-anneal}, we can use either the restricted or non-restricted version of 
%\textsc{GetInput} and \textsc{GetNeighborInput}. In this work, we consider R only, as we have shown in \cite{string-attack-synthesis} that R performs better that NR to synthesize attack efficiently with less number of attack steps.
\vspace{-4mm}
\begin{algorithm}
  \begin{small}
    \caption{\footnotesize \textsc{AttackInput-SA}($C_h, \Psi$) \\
      Generates a low input at each attack step via simulated annealing.
    }
    \label{proc:sim-anneal}
    \begin{algorithmic}[1]
      
      \State $t \leftarrow t_0$, 
      $l_{\mathit{val}} \leftarrow$ \textsc{GetInput}($\Psi, C_h$),
      $\mathcal{I} \leftarrow$ \textsc{MutualInfo}($\Psi, C_h, 
      l_{\mathit{val}}$)
      
      \While{$t \ge t_{\mathit{min}}$}
      
      \State $l_{val} \leftarrow$ \textsc{GetNeighborInput}($l_{\mathit{val}},\Psi, C_h$)
      
      \State $\mathcal{I}_{\mathit{new}} \leftarrow$ \textsc{MutualInfo}($\Psi, 
      C_h, l_{\mathit{val}}$)
      
      \If {$(\mathcal{I}_{\mathit{new}} > \mathcal{I}) \lor \left ({e^{(\mathcal{I}_{\mathit{new}} - 
      \mathcal{I})/t} > \textsc{RandomReal}(0,1)}\right )$}
      
      \State $\mathcal{I} \leftarrow \mathcal{I}_{\mathit{new}}, l^* \leftarrow l_{\mathit{val}}$
           
      \EndIf
      
      \State $t \leftarrow t - (t \times k)$ 
      
      \EndWhile
      
      \State \Return $l^*$
      
    \end{algorithmic}
  \end{small}
\end{algorithm} 
\vspace{-4mm}
\noindent \textit{Searching via Random Model Generation.}
As a base-line search heuristic, we make use of the approach  described above for 
generating low values that are consistent with $C_h$. The simplest approach is to
generate a single random model from $C_l$ and use it as the next attack input.
We call this approach Model-based (M).
A slightly more sophisticated approach is to generate random samples using
$C_l$, compute the expected information gain for each of them using Equation~(\ref{eq:mutinfo}) (i.e.,  objective function is evaluated using the automata-based entropy computation) and then choose the best one. \cite{string-attack-synthesis} evaluates different meta heuristic techniques : genetic algorithm (GA) and simulated annealing (SA) to maximize information leakage and shows that SA performs better than GA. The reason is GA applies mutation and crossover to generate candidate low values. To restrict the search to $l$ values that are consistent with $C_l$, would require implementing mutation and crossover operations with respect to $C_l$. We are not aware of a general approach for doing this, so during GA-based search, mutation and crossover operations can generate low values that are inconsistent with $C_l$ (and hence $C_h$). Note that, such values will have no information
gain and will be ignored during search, but they can slow down the search increasing the search space and hence, we may end up having a higher number of attack steps compared to SA. So, the SA ends up being a more effective meta-heuristic for attack synthesis.

\noindent \textit{Simulated Annealing.}
Simulated annealing (SA) is a meta-heuristic for optimizing an objective
function $g(s)$~\cite{simulatedannealing}.  SA is initialized with a
candidate solution $s_0$. At step $i$, SA chooses a neighbor, $s_i$,
of candidate $s_{i-1}$. If $s_{i}$ is an improvement, i.e., $g(s_i) >
g(s_{i-1})$, then $s_i$ is used as the candidate for the next iteration. If
$s_{i}$ is not an improvement, i.e. $g(s_i) \leq  g(s_{i-1})$, then
$s_i$ is still used as the candidate for the next iteration, but with
a small probability $p$. Intuitively, SA is a controlled random search
that allows a search path to escape from local optima by permitting the
search to sometimes accept worse solutions. The acceptance probability
$p$ decreases exponentially over time, which is modeled using a search
``temperature'' which ``cools off'' and converges to a steady state. Our
use of SA that incorporates automata-based entropy computation is given
in Algorithm~\ref{proc:sim-anneal} where  we
use \textsc{GetNeighborInput} function to get new candidates.

%% file: tex/implementation.tex
%!TEX root = ../main.tex
\vskip -2mm
\paragraph{Implementation.} 
The implementation of our approach consists of two primary components, corresponding to the two main phases described in~\cref{sec:attacker_model}. We implemented Algorithm~\ref{proc:gen-constraints} using Symbolic Path Finder (SPF)~\cite{Pasareanu:2013:ASE}. We implemented Algorithm~\ref{proc:run-attack} as a Java program that takes the observation constraints generated by Algorithm~\ref{proc:gen-constraints} as input, along with $C_h$, $h^*$. \textsc{AttackInput-SA} from~\cref{sec:attack_synthesis_heuristics}
is implemented directly in Java as well. 
We implemented \textsc{GetNeighborInput}, \textsc{ModelCount}, and \textsc{ModelCountIncremental} by extending the existing string model counting tool ABC as described in~\cref{sec:model_counting}. We added these features directly into the C++ source code of ABC along with corresponding Java APIs.

%% file: tex/experiments.tex
\vspace{-12mm}
\input{tables/benchmark-table.tex}
\vspace{6mm}
\noindent \textbf{Benchmark Details.}
To evaluate the effectiveness of our attack synthesis techniques, we experimented on a benchmark of 9 Java canonical programs utilizing various logical and string manipulation operations, setting different sizes and lengths to define the domain of secret value (Table~\ref{tab:string-benchmark}). The functions {\tt{PCI}} and {\tt{PCS}} are password checking implementations. Both compare a user input and secret password but early termination optimization (as described in~\cref{sec:intro}) induces a timing side channel for the first one and the latter is a constant-time implementation. We analyzed the {\tt{SE}} method from the Java String library which is known to contain a timing side channel~\cite{MS14}. We discovered a similar timing side channel in the {\tt{IO}} method from the Java String library. Function {\tt{ED}} is an implementation of the standard dynamic programming algorithm to calculate minimum edit distance of two strings. Function {\tt{CO}} is a basic compression algorithm which collapses repeated substrings within two strings. {\tt{SI}}, {\tt{SCOI}} and {\tt{SCI}} functions check lexicographic inequality ($<, \geq$) of two strings whereas first one directly compares the strings, second one includes \textit{concat} operation with inequality and third one compares characters in the strings.

\noindent \textbf{Experimental Setup.}
%For all experiments, we use a virtual box machine with an Intel Core i7-8750H 2.20 GHz CPU and 16 GB of DDR4 RAM running Ubuntu 18.04.1 LTS, with a Linux 4.15.0-47-generic kernel. We used the Java Hotspot(TM) 64-bit VM, build 1.8.0 191. We ran each experiment for 5 randomly chosen secrets. We present the mean values of the results in Tables~\ref{tab:experiment2}. For SA, we set the temperature range ($t$ to $t_{min}$) from 10 to 0.001 and cooling rate $k$ as 0.1.
For all experiments, we use a desktop machine with an Intel Core i5-2400S 2.50 GHz CPU and 32 GB of DDR3 RAM running Ubuntu 16.04, with a Linux 4.4.0-81 64-bit kernel. We used the OpenJDK 64-bit Java VM, build 1.8.0 171. We ran each experiment for 5 randomly chosen secrets. We present the mean values of the results in Table~\ref{tab:experiment2}. For SA, we set the temperature range ($t$ to $t_{\mathit{min}}$) from 10 to 0.001 and cooling rate $k$ as 0.1.

%\vskip 1mm
\noindent \textbf{Results.}
In this discussion, we describe the quality of a synthesized attack according to these metrics: attack synthesis time, attack length, and overall change in uncertainty about the secret measured as entropy from $\mathcal{H}_{\mathit{init}}$ to $\mathcal{H}_{\mathit{final}}$ and efficiency of incremental attack synthesis in terms of time. Attacks that do not reduce the final entropy to zero are called \textit{incomplete}. Incomplete attacks are mainly due to one of two reasons: the program is not vulnerable to side-channels (for example {\tt{PCS}}) or the observation constraints are very complex, combining lots of path constraints which slows progress too much so that not enough information is leaked within the given time bound (for example {\tt{ED}} and {\tt{SCI}}). For the purpose of direct comparison, in our experiments, we set a bound of 5 hours for SA (slowest technique) on {\tt{ED}} and {\tt{SCI}} and computed a bound for $\mathcal{H}_{\mathit{final}}$ of \textbf{17.28} and \textbf{14.48}, respectively, while all other examples reduced $\mathcal{H}_{\mathit{final}}$ to \textbf{0.0}.These examples are marked with $\mathit{*}$. Note that, M and SA-I techniques can reduce $\mathcal{H}_{\mathit{final}}$ for {\tt{ED}} and {\tt{SCI}} to \textbf{14.34} and \textbf{12.28}, respectively, after one hour.

\textit{Attack Synthesis Time Comparison.} We observe that the model-based technique (M), which only uses $C_l$ to restrict the search space is faster than other techniques, as it greedily uses a random model generated by ABC as the next attack input, with no time required to evaluate the objective function. $M$ quickly generates attacks for most of the functions. We examined those functions and determined that their objective functions are ``flat'' with respect to $l$. Any $l_{\mathit{val}}$ that is a model for $C_l$ at the current step yields the same expected information gain. Figure~\ref{figure:exp1} shows how M can synthesize attacks faster compared to SA (in seconds).\\
\noindent
\begin{minipage}[!h]{0.48\textwidth}
\begin{figure}[H]
   \includegraphics[width=\linewidth]{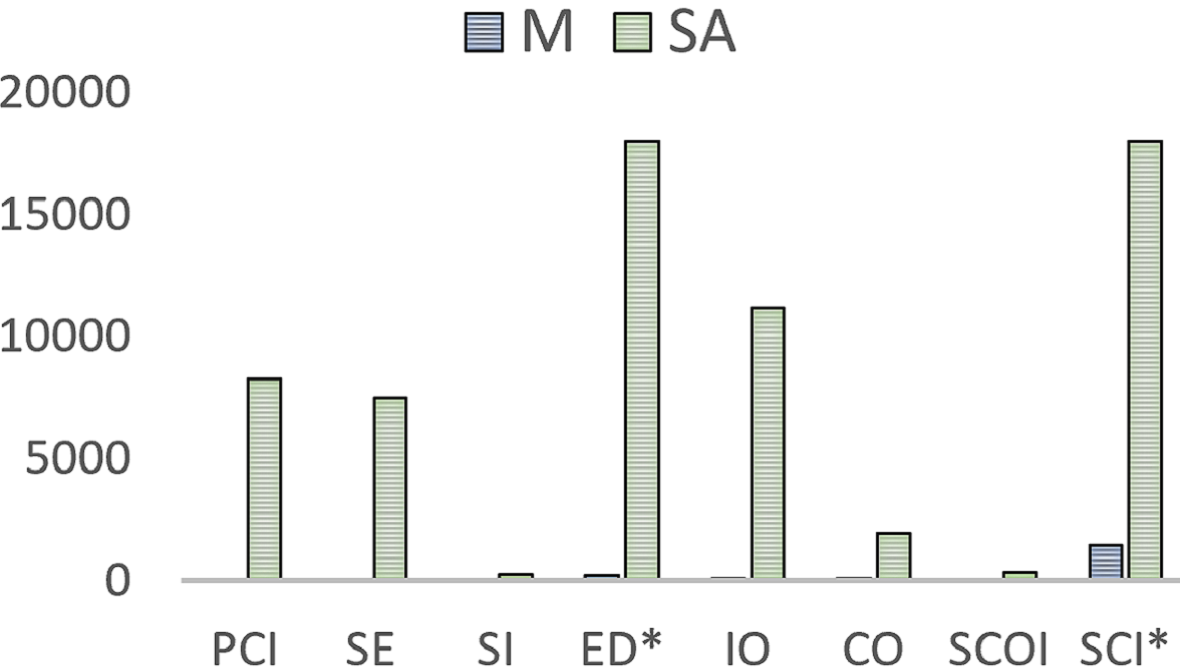}
   \caption{Synthesis Time, M vs SA}
   \label{figure:exp1}
  % \vspace*{-1em}
\end{figure}
\end{minipage}
\hspace{0.04\linewidth}
\begin{minipage}[!h]{0.48\textwidth}
\begin{figure}[H]
   \includegraphics[width=\linewidth]{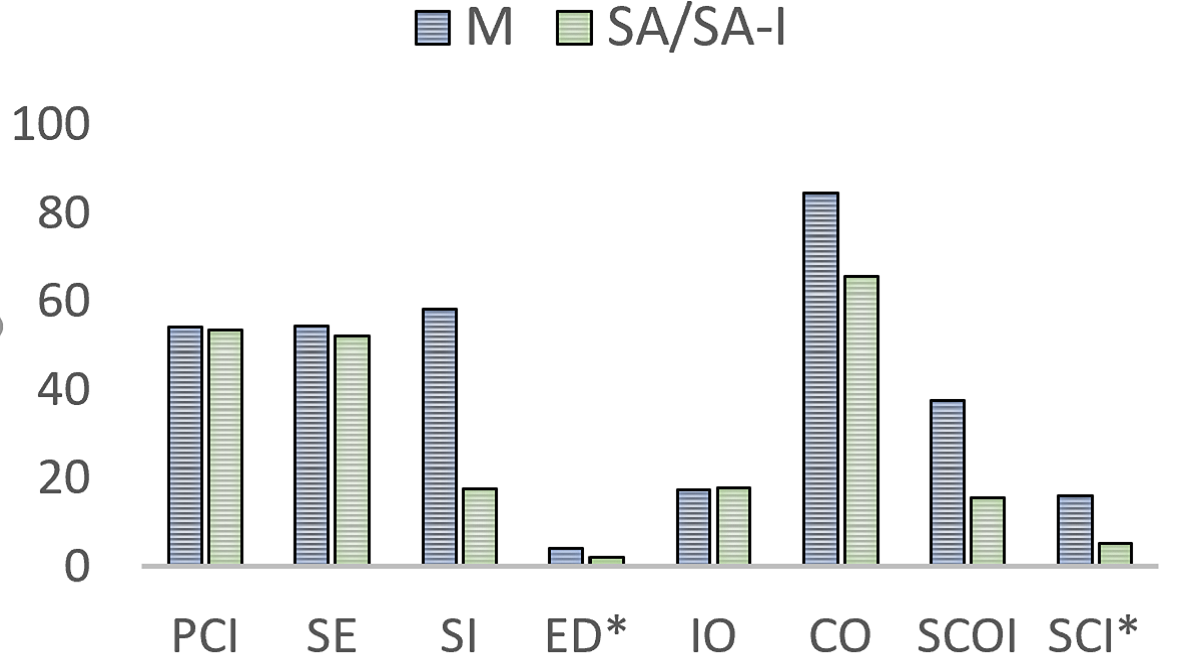}
   \caption{Attack Length, M vs SA/SA-I}
   \label{figure:exp2}
   %\vspace*{-1em}
\end{figure}
\end{minipage}
\vskip 2mm
\textit{Attack Length Comparison.} Although M is fast in synthesizing attacks and generates attacks for each benchmark, experimental results show that it requires more attack steps (in terms of information gain) compared to the attacks generated by meta-heuristic techniques that optimize the objective function. As the experimental results show for the {\tt{SI}}, {\tt{SCOI}} and {\tt{SCI}}, a meta-heuristic technique can reduce $\mathcal{H}_{\mathit{final}}$ further but with fewer attack steps compared to the model-based approach (M). And, this case would be true for any example where different inputs at a specific attack step have different information gain. If attacker is aware of the ``flat" objective function phenomenon, they can proceed with M. In general, M is not efficient to generate an attack with reduced number of attack steps and hence, meta heuristics like SA approach are required. Figure~\ref{figure:exp2} shows how SA is better than M in terms of length of the generated attacks. Note that, we say M vs SA/SA-I as incremental version will make difference in attack synthesis time, not attack length.
\vskip 1mm
\textit{Efficiency of Incremental Attack Synthesis.} On one hand, we can synthesize attacks faster using M but attacks synthesized by M require more attack steps in general. On the other hand, we can synthesize attacks with minimal number of attack steps using SA, but attack synthesis process is slower for SA. Our experiments demonstrate that incremental attack synthesis using SA gives us fast attack synthesis without increasing the attack length. We compare incremental version of SA (SA-I) against SA. Figure~\ref{figure:exp3} shows SA-I is an order of magnitude faster than SA for all the examples from the benchmark. We also compare SA-I against M and Figure~\ref{figure:exp4} shows that SA-I is comparable to M in terms of attack synthesis time (in seconds). \\
\noindent
\begin{minipage}[!h]{0.48\textwidth}
\begin{figure}[H]
   \includegraphics[width=\linewidth]{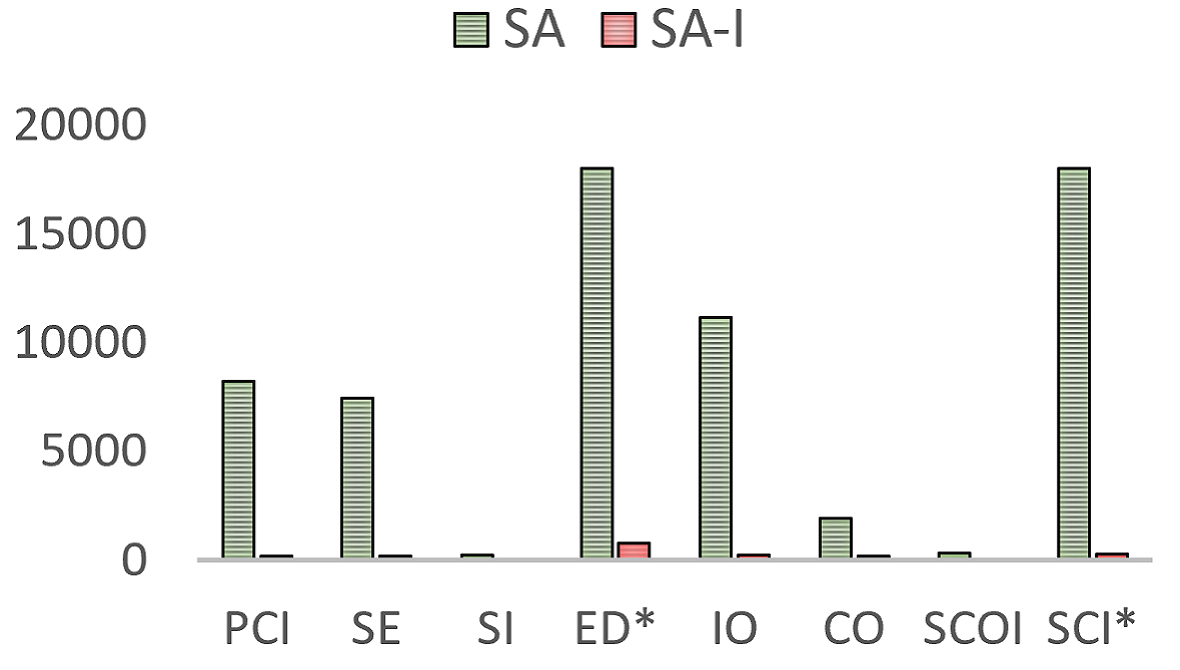}
   \caption{Synthesis Time, SA vs SA-I}
   \label{figure:exp3}
   %\vspace*{-1em}
\end{figure}
\end{minipage}
\hspace{0.04\linewidth}
\begin{minipage}[!h]{0.48\textwidth}
\begin{figure}[H]
   \includegraphics[width=\linewidth]{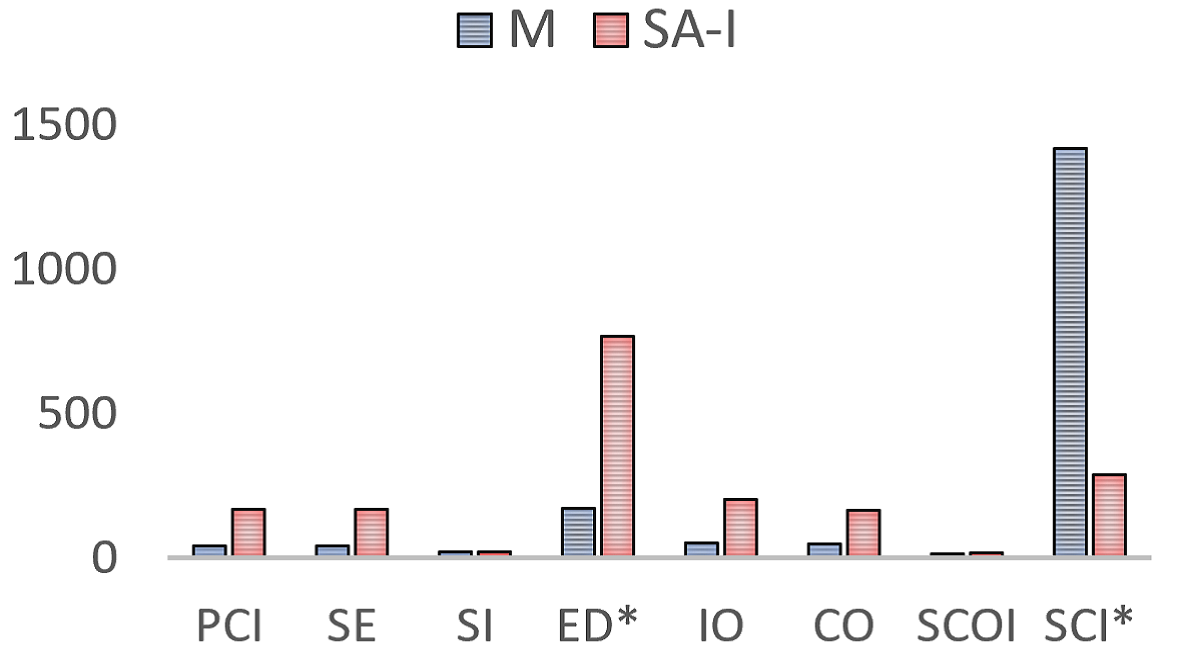}
   \caption{Synthesis Time, M vs SA-I}
   \label{figure:exp4}
   %\vspace*{-1em}
\end{figure}
\end{minipage}
\input{tables/experiment-table.tex}
\vskip -10mm
\textit{Vulnerability to Side-Channels.} Finally, we observe that some of our selected benchmarks are more secure against our attack synthesizer than others. In particular, {\tt{PCS}}, a constant-time implementation of password checking, did not leak any information through the side channel. One of the examples from the benchmark, {\tt{ED}} also did not succumb to our approach easily, due to the relatively large number of generated constraints (2170), indicating a much more complex relationship between the inputs and observations. To summarize, our experiments indicate that our attack synthesis approach is able to construct side-channel attacks, providing evidence of vulnerability (e.g. {\tt{PCI}}). Further, when attack synthesizer fails to generate attacks ({\tt{PCS}}), or is only able to extract a relatively small information after many steps of significant computation time ({\tt{ED}}), it provides evidence that the function under test is comparatively safer against side-channel attacks. Table~\ref{tab:experiment2} shows results for {\tt{PCS}}, hardly reducing $\mathcal{H}_{\mathit{final}}$ even after running for 1 hour for M, SA and SA-I.

%% file: tables/benchmark-table.tex
\renewcommand{\floatpagefraction}{.9}
\vskip 4mm
\begin{table}[!h]
\centering
\caption{Benchmark details with the number of path constraints ($|\Phi|$) and the 
number of merged observation constraints ($|\Psi|$).}
\vskip 2mm
\label{tab:string-benchmark}
\scalebox{1.0}{
\begin{scriptsize}
\begin{tabular}{|l|c|c|c|c|r|r|}
\hline

Benchmark & ID & Operations & {\makecell{Low\\Length}} & {\makecell{High\\Length}} & 
$|\Phi|$ 
& $|\Psi|$\\ \hline \hline

\tt{passCheckInsec} & PCI &
{\makecell{charAt,length}} & 4 & 4 & 5 & 5 \\
\hline 
\tt{passCheckSec} & PCS & {\makecell{charAt,length}} & 4 & 4 & 16 & 1 \\ \hline 

\tt{stringEquals} & SE & {\makecell{charAt,length}} & 4 & 4 & 9 & 9 
\\ 
\hline 

\tt{stringInequality} & SI & 
{\makecell{$<$,$\ge$}} & 4 & 4 & 2 & 2 \\
\hline

\tt{stringConcatInequality} & SCOI & {\makecell{concat,$<$,$\ge$}} & 4 & 4 & 
2 & 2 \\ \hline

\tt{stringCharInequality} & SCI & {\makecell{charAt,length,$<$,$\ge$}} & 4 & 4 & 
80 & 2 \\ \hline

\tt{indexOf} & IO & {\makecell{charAt,length}} & 1 & 8 & 9 & 9 \\ \hline 

\tt{compress} & CO & {\makecell{begins,substring,length}} & 4 & 4 & 5 & 
5 \\ \hline 

\tt{editDistance} & ED & {\makecell{charAt,length}} & 4 & 4 & 2170 & 22 \\ 
\hline

\end{tabular}
\end{scriptsize}
}
\end{table}
\vspace{-12mm}

%% file: tables/experiment-table.tex
\renewcommand{\floatpagefraction}{.9}

\vspace{-8mm}
\begin{table}[htbp]
  \caption{Experimental results for secure password checker ({\tt{PCS}}). Time bound is set as 3600 seconds.}
  \vspace{-4mm}
  \begin{center}
  	\scalebox{1.0}{
    \begin{scriptsize}
      \begin{tabular}{|l|r|l|r|r|r|}
        \hline
        \textbf{ID} & \textbf{$\mathcal{H}_{\mathit{init}}$} &
        \textbf{Metrics} & \textbf{M} & \textbf{SA} & \textbf{SA-I} \\
        \hline \hline
        \multirow{2}{*}{\tt{PCS}} &
        \multirow{2}{*}{18.8} &
        Steps & 108 & 14 & 99 \\
        \cline{3-6} 
        & & $\mathcal{H}_{\mathit{final}}$ & 18.8 & 18.8 & 18.8 \\
        \cline{3-6} 
        \hline
      \end{tabular}
    \end{scriptsize}
	}
    \label{tab:experiment2}
  \end{center}
\end{table}

%% file: tex/case-studies.tex
%!TEX root = ../main.tex
\vskip -2mm
Our experimental results show that synthesizing attacks face scalability issues for programs leading to large numbers of complex observation constraints. Note that, this limitation depends on the limitations of the building blocks: constraint solvers and model counters. The more powerful these tools become, more powerful attack synthesis will be. We now present two case studies.
\vskip 1mm
\noindent \textit{CRIME Attack.} The “Compression Ratio Info-leak Made Easy” (CRIME) attack~\cite{CRIMEattack} allows an attacker to learn fields of encrypted web session headers by injecting extraneous text ($l$) into a procedure that compresses and encrypts the header ($h$). Despite the encryption, an attacker can infer how much of the injected text matched the unknown header by observing the number of bytes in the compressed result~\cite{BAP16, CRIMEattack}. Our approach automatically synthesizes this attack. Symbolic execution of the compression function (LZ77T) for a secret of length 3 and alphabet size 4 yields 187 path constraints and 4 observations, leading to 4 observation constraints. M synthesizes an attack in 6.8 steps within 468.5 seconds. SA-I could generate the attack in 7.8 steps within 757.4 seconds. SA-I does not improve over M due to ``flat'' objective function. Note that~\cite{BAP16} performs leakage quantification for this example but does not synthesize attacks.

\vskip 1mm
\noindent \textit{Law Enforcement Database.}
The Law Enforcement Employment Database Server is a network service application included as a part of the DARPA STAC program~\cite{darpa,stac}. This application provides access to records about law enforcement personnel. Employee information is determined by an employee ID number. The database contains restricted and unrestricted employee information. Users can search ranges of employee IDs. If an ID query range contains one or more restricted IDs, the returned data will not contain the restricted IDs. We decompiled the application and then symbolically executed the \texttt{channelRead0} method from the \texttt{UDPServerHandler} class which performs the database search operation. We limited the domain of ids to 1024, added 30 unrestricted IDs and 1 restricted ID. Symbolic execution gives us 1669 path constraints with 162 distinguishable observations  ($\delta = 10$ instructions). M generates attack with an attack length of 8.2 in 270.1 seconds whereas SA-I generates an attack with length of 6.5 in 810.7 seconds. SA-I requires less attack length as the objective function is not ``flat''.

%% file: tex/related.tex
%!TEX root = ../main.tex
\vskip -2mm
There are two previous results that are most closely related to our work~\cite{BAP16,PhanCSF2017}. The first focuses on quantifying information leakage through side channels for programs manipulating strings~\cite{BAP16}. This work assumes that the given program has a segment oracle side-channel vulnerability and then quantifies the amount of information leakage for that vulnerability. Other recent work synthesizes side-channel attacks using either entropy-based or SAT-based objective functions, but works only for linear arithmetic and bit-vector constraints~\cite{PhanCSF2017}, using model counters and constraint solvers for those theories~\cite{DeLoera20041273}. This earlier approach also relies on generation of a closed form objective function that represents the information leakage, and uses model counting techniques that specialize on linear integer arithmetic to construct such a function. In contrast, our approach is more general and can handle any program with numeric, string and mixed constraints. Furthermore, our approach does not require a closed form solution for the objective function as we use meta heuristics to search for input values that leak maximum information. Both of these earlier approaches use constraint solving and model counting queries to quantify the information leakage, but they do not use an incremental approach and, therefore, re-compute many sub-queries.

There are many works on analyzing side-channels in various settings~\cite{Brumley:2003:RTA:1251353.1251354,
Chen:2010:SLW:1849417.1849974, DBLP:conf/sp/MardzielAHC14,
Do2015, Pasareanu:2016:MSC,BAP16}. A few recent works address either synthesizing attacks or quantifying information leakage under a model where the attacker can make multiple invocations of the system~\cite{DBLP:conf/ccs/KopfB07,DBLP:conf/sp/MardzielAHC14,Pasareanu:2016:MSC,BAP16,BRB18}.
Single-run analysis is addressed in \cite{Heusser:2010:QIL:1920261.1920300}
where bounded model checking is used over the $k$-composition of a program to determine if it can yield $k$ different outputs. Further, LeakWatch~\cite{DBLP:conf/esorics/ChothiaKN14} estimates leakage in Java programs based on sampling program executions on concrete inputs. There has been work on multi-run analysis using enumerative techniques~\cite{DBLP:conf/ccs/KopfB07}. None of these earlier results present a symbolic and incremental approach to adaptive attack synthesis as we present in this paper. 

%There is a growing body of work on string analysis~\cite{KGG09,ZZG13,LG13,AAC14,TCJ14}; however none of these approaches provide model-counting functionality. 

%There are three existing tools for model counting string constraints: SMC~\cite{LSS14}, \STS~\cite{TCJ17}, and ABC~\cite{ABB15}. 

Due to the importance of model counting in quantitative program analyses, model counting constraint solvers are gaining increasing attention~\cite{LSS14,TCJ17,ABB15}.
ABC is the only one that supports string, numeric and mixed constraints. We extended ABC to perform incremental model counting for our attack synthesis approach. Other work in quantifying information
leakage~\cite{PhanCSF2017,Pasareanu:2016:MSC,Phan:2014:QILURA,BRB18} have used symbolic execution and model-counting techniques for linear integer arithmetic.

Preliminary results from this paper were discussed in a short workshop paper~\cite{string-attack-synthesis} which did not include our results on incremental attack synthesis and real world scenarios but was focused on studying different meta-heuristic techniques such as random search, genetic algorithm, and simulated annealing.

%% file: tex/conclusion.tex
\vskip -2mm
In this paper, we presented techniques for synthesizing adaptive side-channel attacks on programs that manipulate string and numeric values. Our approach uses meta-heuristics for selecting public inputs that maximize the amount of information gained about the secret by computing the amount of remaining uncertainty about the secret using entropy and model counting. We exploit the iterative nature of attack synthesis by presenting an incremental approach that reuses results from prior iterations. We implemented our attack synthesis approach for Java programs using symbolic execution tool SPF and ABC model counter. We evaluated the effectiveness of our attack synthesis approach on several functions and two case studies. Our experiments demonstrate that our incremental approach greatly improves the efficiency of attack synthesis.

%we observed that adaptive attack
%synthesis that is based on constraint solving and model counting should
%take into account the inherent incremental nature of constraints which
%characterize the secret. Previous works in this area did not exploit this
%feature of the generated constraints. We implemented incremental string
%constraint solving and model counting on top of an existing automata-based
%model counter, ABC. 

%% file: main.bbl
\begin{thebibliography}{10}

\bibitem{stac}
\url{https://github.com/Apogee-Research/STAC/tree/master/Engagement_Challenges}.

\bibitem{darpa}
Space/time analysis for cybersecurity (stac).
\newblock
  \url{https://www.darpa.mil/program/space-time-analysis-for-cybersecurity}.

\bibitem{ABB15}
Abdulbaki Aydin, Lucas Bang, and Tevfik Bultan.
\newblock Automata-based model counting for string constraints.
\newblock In {\em Proceedings of the 27th International Conference on Computer
  Aided Verification (CAV)}, pages 255--272, 2015.

\bibitem{aydin2018parameterized}
Abdulbaki Aydin, William Eiers, Lucas Bang, Tegan Brennan, Miroslav Gavrilov,
  Tevfik Bultan, and Fang Yu.
\newblock Parameterized model counting for string and numeric constraints.
\newblock In {\em Proceedings of the 2018 26th ACM Joint Meeting on European
  Software Engineering Conference and Symposium on the Foundations of Software
  Engineering}, pages 400--410. ACM, 2018.

\bibitem{BAP16}
Lucas Bang, Abdulbaki Aydin, Quoc-Sang Phan, Corina~S. Pasareanu, and Tevfik
  Bultan.
\newblock String analysis for side channels with segmented oracles.
\newblock In {\em Proceedings of the 24th ACM SIGSOFT International Symposium
  on the Foundations of Software Engineering}, 2016.

\bibitem{BRB18}
Lucas Bang, Nicolas Rosner, and Tevfik Bultan.
\newblock Online synthesis of adaptive side-channel attacks based on noisy
  observations.
\newblock In {\em Proceedings of the IEEE European Symposium on Security and
  Privacy (EuroS\&P)}, 2018.

\bibitem{Brumley:2003:RTA:1251353.1251354}
David Brumley and Dan Boneh.
\newblock {Remote Timing Attacks Are Practical}.
\newblock In {\em Proceedings of the 12th Conference on USENIX Security
  Symposium - Volume 12}, SSYM'03, pages 1--1, Berkeley, CA, USA, 2003. USENIX
  Association.

\bibitem{Chen:2010:SLW:1849417.1849974}
Shuo Chen, Rui Wang, XiaoFeng Wang, and Kehuan Zhang.
\newblock Side-channel leaks in web applications: A reality today, a challenge
  tomorrow.
\newblock In {\em Proceedings of the 2010 IEEE Symposium on Security and
  Privacy}, SP '10, pages 191--206, Washington, DC, USA, 2010. IEEE Computer
  Society.

\bibitem{DBLP:conf/esorics/ChothiaKN14}
Tom Chothia, Yusuke Kawamoto, and Chris Novakovic.
\newblock Leakwatch: Estimating information leakage from java programs.
\newblock In {\em Computer Security - {ESORICS} 2014 - 19th European Symposium
  on Research in Computer Security, Wroclaw, Poland, September 7-11, 2014.
  Proceedings, Part {II}}, pages 219--236, 2014.

\bibitem{Cover2006}
Thomas~M. Cover and Joy~A. Thomas.
\newblock {\em Elements of Information Theory (Wiley Series in
  Telecommunications and Signal Processing)}.
\newblock Wiley-Interscience, 2006.

\bibitem{MS14}
Joel~Sandin Daniel~Mayer.
\newblock Time trial: Racing towards practical remote timing attacks.
\newblock
  \url{https://www.nccgroup.trust/globalassets/our-research/us/whitepapers/TimeTrial.pdf},
  2014.

\bibitem{Do2015}
Quoc~Huy Do, Richard Bubel, and Reiner H{\"a}hnle.
\newblock {Exploit Generation for Information Flow Leaks in Object-Oriented
  Programs}.
\newblock In {\em ICT Systems Security and Privacy Protection: 30th IFIP TC 11
  International Conference, SEC 2015, Hamburg, Germany, May 26-28, 2015,
  Proceedings}, pages 401--415, Cham, 2015. Springer International Publishing.

\bibitem{Heusser:2010:QIL:1920261.1920300}
Jonathan Heusser and Pasquale Malacaria.
\newblock {Quantifying information leaks in software}.
\newblock In {\em Proceedings of the 26th Annual Computer Security Applications
  Conference}, ACSAC '10, pages 261--269, New York, NY, USA, 2010. ACM.

\bibitem{Kel02}
John Kelsey.
\newblock Compression and information leakage of plaintext.
\newblock In {\em Fast Software Encryption, 9th International Workshop, {FSE}
  2002, Leuven, Belgium, February 4-6, 2002, Revised Papers}, pages 263--276,
  2002.

\bibitem{King:1976:SEP:360248.360252}
James~C. King.
\newblock {Symbolic execution and program testing}.
\newblock {\em Commun. ACM}, 19(7):385--394, July 1976.

\bibitem{simulatedannealing}
Scott Kirkpatrick, C~Daniel Gelatt, and Mario~P Vecchi.
\newblock Optimization by simulated annealing.
\newblock {\em science}, 220(4598):671--680, 1983.

\bibitem{DBLP:conf/ccs/KopfB07}
Boris K{\"{o}}pf and David~A. Basin.
\newblock An information-theoretic model for adaptive side-channel attacks.
\newblock In Peng Ning, Sabrina De~Capitani di~Vimercati, and Paul~F. Syverson,
  editors, {\em Proceedings of the 2007 {ACM} Conference on Computer and
  Communications Security, {CCS} 2007}, pages 286--296. {ACM}, 2007.

\bibitem{Law09.2}
Nate Lawson.
\newblock Timing attack in google keyczar library.
\newblock
  \url{https://rdist.root.org/2009/05/28/timing-attack-in-google-keyczar-library/},
  2009.

\bibitem{DeLoera20041273}
Jesús A.~De Loera, Raymond Hemmecke, Jeremiah Tauzer, and Ruriko Yoshida.
\newblock Effective lattice point counting in rational convex polytopes.
\newblock {\em Journal of Symbolic Computation}, 2004.
\newblock Symbolic Computation in Algebra and Geometry.

\bibitem{LSS14}
Loi Luu, Shweta Shinde, Prateek Saxena, and Brian Demsky.
\newblock A model counter for constraints over unbounded strings.
\newblock In {\em Proceedings of the {ACM} {SIGPLAN} Conference on Programming
  Language Design and Implementation}, page~57, 2014.

\bibitem{DBLP:conf/sp/MardzielAHC14}
Piotr Mardziel, M{\'{a}}rio~S. Alvim, Michael~W. Hicks, and Michael~R.
  Clarkson.
\newblock Quantifying information flow for dynamic secrets.
\newblock In {\em 2014 {IEEE} Symposium on Security and Privacy, {SP} 2014,
  Berkeley, CA, USA, May 18-21, 2014}, pages 540--555, 2014.

\bibitem{Nel10}
Taylor Nelson.
\newblock Widespread timing vulnerabilities in openid implementations.
\newblock
  \url{http://lists.openid.net/pipermail/openid-security/2010-July/001156.html},
  2010.

\bibitem{PhanCSF2017}
Quoc{-}Sang Phan, Lucas Bang, Corina~S. Pasareanu, Pasquale Malacaria, and
  Tevfik Bultan.
\newblock Synthesis of adaptive side-channel attacks.
\newblock In {\em 30th {IEEE} Computer Security Foundations Symposium, {CSF}
  2017, Santa Barbara, CA, USA}, 2017.

\bibitem{Phan:2014:QILURA}
Quoc-Sang Phan, Pasquale Malacaria, Corina~S. P\u{a}s\u{a}reanu, and Marcelo
  d'Amorim.
\newblock {Quantifying Information Leaks Using Reliability Analysis}.
\newblock In {\em Proceedings of the 2014 International SPIN Symposium on Model
  Checking of Software}, SPIN 2014, pages 105--108, New York, NY, USA, 2014.
  ACM.

\bibitem{Pasareanu:2016:MSC}
Corina~S. P\u{a}s\u{a}reanu, Quoc-Sang Phan, and Pasquale Malacaria.
\newblock {Multi-run side-channel analysis using Symbolic Execution and
  Max-SMT}.
\newblock In {\em Proceedings of the 2016 IEEE 29th Computer Security
  Foundations Symposium}, CSF '16, Washington, DC, USA, 2016. IEEE Computer
  Society.

\bibitem{Pasareanu:2013:ASE}
Corina~S. P\u{a}s\u{a}reanu, Willem Visser, David Bushnell, Jaco Geldenhuys,
  Peter Mehlitz, and Neha Rungta.
\newblock {Symbolic PathFinder: integrating symbolic execution with model
  checking for Java bytecode analysis}.
\newblock {\em Automated Software Engineering}, pages 1--35, 2013.

\bibitem{CRIMEattack}
J.~Rizzo and T.~Duong.
\newblock The crime attack.
\newblock Ekoparty Security Conference, 2012.

\bibitem{string-attack-synthesis}
Seemanta Saha, Ismet~Burak Kadron, William Eiers, Lucas Bang, and Tevfik
  Bultan.
\newblock Attack synthesis for strings using meta-heuristics.
\newblock {\em SIGSOFT Softw. Eng. Notes}, 43(4):56--56, January 2019.

\bibitem{shannon48}
Claude Shannon.
\newblock A mathematical theory of communication.
\newblock {\em Bell System Technical Journal}, 27:379--423, 623--656, July,
  October 1948.

\bibitem{Smi09}
Geoffrey Smith.
\newblock On the foundations of quantitative information flow.
\newblock In {\em Proceedings of the 12th International Conference on
  Foundations of Software Science and Computational Structures (FOSSACS)},
  pages 288--302, 2009.

\bibitem{TCJ17}
Minh-Thai Trinh, Duc-Hiep Chu, and Joxan Jaffar.
\newblock Model counting for recursively-defined strings.
\newblock In {\em Computer Aided Verification - 29th International Conference,
  {CAV} 2017, Heidelberg, Germany, Proceedings, Part {II}}, pages 399--418,
  2017.

\end{thebibliography}
